# Exploration of vitrification of water and Kauzmann entropy through complex specific heat: A journey through 'No Man's Land'

**Shinji Saito**[1,*] **and Biman Bagchi**[2,*]

[1] Institute for Molecular Science, Myodaiji, Okazaki, Aichi, 444-8585, Japan and The Graduate University for Advanced Studies, Myodaiji, Okazaki, Aichi, 444-8585, Japan.

[2] Indian Institute of Science, Bangalore, 560012, India.

Frequency dependent specific heat, introduced by Grest and Nagel, offers valuable insight into the vitrification of supercooled liquid. We calculate this quantity and other thermodynamic properties of supercooled liquid water by varying temperature and density across the "no man's land" all the way to the formation of amorphous ice. The calculations are aided by very long computer simulations, often more than 50 $\mu$s long. Density fluctuations that arise from the proximity to a putative liquid-liquid (LL) transition at 228 K, cast a long shadow on the properties of water, both above and below the LL transition. We carry out the calculation of the quantum mechanical static and frequency-dependent specific heats by combining seminal works by Lebowitz, Percus, and Verlet and Grest and Nagel with the harmonic approximation for the density of states. The obtained values are in quantitative agreement with all available experimental and numerical results of specific heats for both supercooled water and ice. We calculate the entropy at all the state points by integrating the specific heat. We find that the quantum corrected-contributions of intermolecular vibrational entropy dominate the excess entropy of amorphous phases over the crystal over a wide range of temperature. Interestingly, the vibrational entropy lowers the Kauzmann temperature, $T_K$, to 130 K, just below the experimental glass-to-liquid water transition temperature, $T_g$, of 136 K and the calculated $T_g$ of 135 K in our previous study. A straightforward extrapolation of high temperature entropy from 250 K to below however would give a much higher value of $T_K \sim 190$ K. The calculation of Lindemann ratios places the melting of amorphous ice ~135 K. The amorphous state exhibits an extremely short correlation length for the distance dependence of orientational correlation.

shinji@ims.ac.jp
bbagchi@iisc.ac.in, profbiman@gmail.com

## I. INTRODUCTION

Formation of amorphous ice is still shrouded in mystery despite the fact that it is the predominant form of water in the universe outside planet earth, in asteroids and meteoroids, and also in the moons of the planets, such as Jupiter. Yet, we know little about the mechanism of its formation. Supercooled water undergoes a glass-to-liquid transition at 136 K. The liquid water so formed is highly viscous, and although highly metastable, remains in the liquid state till ~150 K when it undergoes crystallization to cubic ice. The water glass transition could be a phenomenon of astrophysical significance, yet its nature and that of the glassy water has remained largely incomplete. Due to the uniqueness of water, the formation of amorphous can also provide valuable insight into the stability of glasses towards melting and re-crystallization – two phenomena of practical importance in glasses formed from molecular liquids where intermolecular vibrations are important.

Much of the theoretical work on liquid-glass transition originates from the landmark research of Kauzmann who proposed that a supercooled liquid would inevitably undergo a glass transition at or before the temperature where the extrapolated entropy of the liquid would become equal to the entropy of the crystal.[1] This temperature, widely known as the Kauzmann temperature $T_K$, is defined by

$$\Delta S(T_K) \equiv S_{liq}(T_K) - S_{cry}(T_K) = 0 . \tag{1}$$

One indeed finds that for a number of fragile glass-forming liquids, the experimentally determined (by extrapolation of entropy) Kauzmann temperatures are placed slightly below $T_g$. *There is yet no (theoretical or experimental) quantitative estimate of $T_K$ for water*. In fact, few calculations exist for the Kauzmann temperature for real liquids.

The entropy crisis scenario by Kauzmann has over the years motivated a large number of theoretical investigations in the area.[2-4] The well-known theory of Gibbs and DiMarzio assigns an ideal glass transition at the temperature $T_2$ where the configuration entropy of the liquid goes to zero.[5, 6] The Gibbs-DiMarzio condition of glass transition *bypasses the role of vibrational entropy in the observed glass transition*. Since a glass transition is regarded as a kinetic phenomenon characterized by a rapid growth of viscosity, a simultaneous rapid decrease in entropy can provide a useful thermodynamic view of the phenomenon.[7, 8] However, this issue has not been addressed adequately by examining both the configurational and vibrational entropies of liquid and crystal separately. Such a calculation is difficult especially for water because of a large number of delocalized intermolecular modes due to extensive hydrogen bonding. This can be seen in Fig. 1 (see below) where the significant difference between the density of states (DOSs) of amorphous water and crystalline ice is presented.

In our recent paper, we offered a unified picture of anomalous properties of water between 270 K and 130 K.[9] The study concluded that the vitrification of water occurs at least in three distinct stages, with two crossovers in thermodynamic and dynamic properties. First, the dynamical transition at 220 K was identified with the crossing of the Widom line. A second transition at 190 K was found to occur by the fragmentation and isolation of the locally distorted-structured clusters. The locally distorted-structured clusters corresponding to high-density liquid (HDL) region contain significant number of non-tetrahedral coordinated water molecules. These distorted-structured molecules dominate the structural, dynamic, and thermodynamic properties of liquid water above 230 K. Below 190 K, where very long simulations were needed to obtain meaningful converged results, these locally distorted-structured molecules were found to exist as the dispersed three- and five-coordinated defects in the locally tetrahedral-structured network, that is the network formed by low-density liquid (LDL).

In the same work, a detailed theoretical analysis of the defects was presented.[9] It was found that one particular type of three-coordinated water molecules (denoted by $H^1O^2$ to indicate one hydrogen bond to one of the hydrogen atoms and two hydrogen bonds to the oxygen atom, i.e. one HB donor and two HB acceptors) plays an important role in the structural relaxation of extensive HB network at temperatures below 190 K. The study further established, for the first time, these defects, scattered throughout the continuous HB network and hard to annihilate, are responsible for the significant reduction of the glass transition temperature of water. The said paper proposed that the fragmentation and isolation of locally distorted-structured clusters at 190 K and the hitherto unknown role of the active $H^1O^2$-defect together serve to explain many of the anomalies in deeply supercooled region of liquid water, such as the limited existence of the observed supercooled liquid state between 136 K and 150 K. The detailed enumeration of the role of the defects indeed provided a deeper understanding of the myriad of anomalous properties of supercooled water than was available before.

The objective of the present work can be articulated as follows. Here, we investigate a thermodynamic perspective of the water glass transition phenomenon. Such a perspective has been missing so far. We inquire how freezing or glassification proceeds and how that is reflected in the slow-down of the vibrational and configurational motions. The evaluation of thermodynamic properties, especially entropy, can provide an estimate of the Kauzmann temperature.

As mentioned above, supercooled water undergoes two dynamical transitions at 220 K and 190 K. Although there is no phase transition at the ambient conditions, very slow fluctuations accompany a crossover from an HDL- to an LDL-dominated phase as the temperature is lowered, especially below 220 K. The presence of these slow fluctuations makes the calculation challenging and thus long molecular simulations, more than 50 $\mu$s, are required to capture the fluctuations. In addition, we need to have a special care for quantum effects for water and ice which have intermolecular vibrational motions from 50 cm$^{-1}$ to 1000 cm$^{-1}$ since we are interested in low temperature behaviors. Furthermore, the thermodynamic properties of deeply supercooled water can be daunting because of the simultaneous presence of vibrational and configurational dynamics which are of course strongly coupled.

Careful calculations performed here (and detailed below) lead to several novel results on thermodynamic properties related to the water glass transition phenomenon. We first obtain the quantum-corrected specific heat and entropies of liquid water and ice over a wide range of temperatures based on the seminal studies by Lebowitz, Percus, and Verlet and Grest and Nagel. The frequency-dependent specific heat and entropy developed in this study make it possible to reveal how the vibrational and conformational dynamics proceed toward vitrification as the temperature is lowered. Furthermore, we could make a prediction of the Kauzmann temperature, $T_K$, of water to be 130 K, which is just below the calculated glass transition temperature, 135 K. We find an unexpected large vibrational contribution to the determination of $T_K$ and to the specific heat and entropy of deeply supercooled water below 190 K.

## II. COMPUTATIONAL DETAILS

The details of simulations are found in our previous study.[9] Here, we summarize only the brief explanation. We have performed molecular dynamics (MD) simulations with 1000 water molecules under both NVE and NPT conditions using our own program and GROMACS.[10] The TIP4P/2005 model potential[11] was used for the water molecules. In these simulations, the periodic boundary condition was employed and the long-range electric interactions were calculated by using the Ewald sum. The velocity Verlet method with a time step of 2 fs was used to solve the equation of motions.

We determined the densities at 22 temperatures from 300 K to 130 K, after the enough long equilibration run with NPT simulations. Several independent long production runs of NVE simulations were performed to examine the dynamical properties after the equilibration runs at individual temperatures: three sets of production runs at each temperature for 300 K – 190 K, and five sets of 80 $\mu$s-production runs at 180 K and six sets at each temperature lower below 180 K. As a result, we have performed the totally 45 ns-, 30 $\mu$s-, 400 $\mu$s-, and 480 $\mu$s-production runs at 300, 190, 180 K, and for 170 K – 145 K, respectively. For ice Ih, we carried out NVE MD simulations at 275, 250, 200, 150, 130, 100, 50, 25, 15, 12, 10, and 7 K at the density determined with the TIP4P/2005 model by Noya et al.[12] For each temperature, two independent 15 ns simulations were performed.

In addition to the MD simulations, quenched normal modes (QNM) and instantaneous normal mode (INMs) analyses have been performed to examine the characters of motions and to make quantum corrections for the specific heats of ambient, supercooled liquid and amorphous water and ice Ih. Quenched structures are the energy local minima along the trajectories obtained from MD simulations. A conjugated gradient method has been used to obtain the quenched structures. QNMs and INMs have been calculated at more than 100 configurations at well separated intervals depending on temperature.

## III. THEORETICAL ANALYSIS OF STRUCTURE AND DYNAMICS OF SUPERCOOLED WATER AND ICE

### A. Density of states of quenched and instantaneous structures of water and ice

**Figures 1 (a)** and **1 (b)** show the DOSs of QNM of (amorphous) liquid water and ice Ih at 250 K and 130 K, respectively. Below 350 cm$^{-1}$, there are two peaks at ~60 cm$^{-1}$ and ~280 cm$^{-1}$ arising from the intermolecular hydrogen bond (HB) bending and HB stretching motions. The DOS of HB stretching motion in liquid and amorphous water is lower than that of HB bending motion, whereas it is sharp and high in ice. Above 400 cm$^{-1}$, on the other hand, there is a broad band arising from the libration motion, particularly in liquid (amorphous) water. With decreasing temperature, the peak of HB stretching motion increases, and shifts to higher frequency in both water and ice. A similar blue shift is seen in the libration motion and thus the gap between the translational and libration bands increases with lowering temperature. The difference between the peaks of the intermolecular motions between water and ice can be found in the difference of entropy between water and ice: the blue shift of the low frequency edge of libration motion in ice can be seen as a positive contribution in the entropy difference between liquid and ice [Fig. 8 (c)]. We also note that the DOS at ~30 cm$^{-1}$ of liquid is also large compared with that of ice [as the results shown in Fig. 8 (d)], indicating the presence of Boson peak.

We calculated the scaled inverse participation ratio (IPR) of mode $n$ defined by[13]

$$IPR_n = \frac{1}{N}\left[\sum_{i=1}^{N}\left(\sum_{\alpha=1}^{6}(U_{i\alpha,n})^2\right)^2\right]^{-1}, \qquad (2)$$

where $U_{i\alpha,n}$ is element $\alpha$ ($\alpha$=1, 2, ⋅, ⋅, 6) corresponding to three translational and three rotational motions) of molecule $i$ of normal mode $n$ in the system with $N$ molecules. The scaled IPR would be $1/N$, when a mode is localized at a single molecule, whereas it would be 1 when a mode is completely delocalized to all the molecules. **Figures 1 (c)** and **1 (d)** show the IPRs at 250 K and 130 K, respectively. By comparing the IPRs of liquid water and ice Ih, the modes of ice are found to be generally more delocalized, especially, the HB stretch and libration motions. In addition, the intermolecular translational motion below 350 cm$^{-1}$ is more delocalized than the libration motions above 400 cm$^{-1}$. As seen in the DOSs, the IPR of ice does not show any significant temperature dependence. On the other hand, the modes in liquid water become slightly delocalized with decreasing temperature, especially the intermolecular translational motion below 300 cm$^{-1}$.

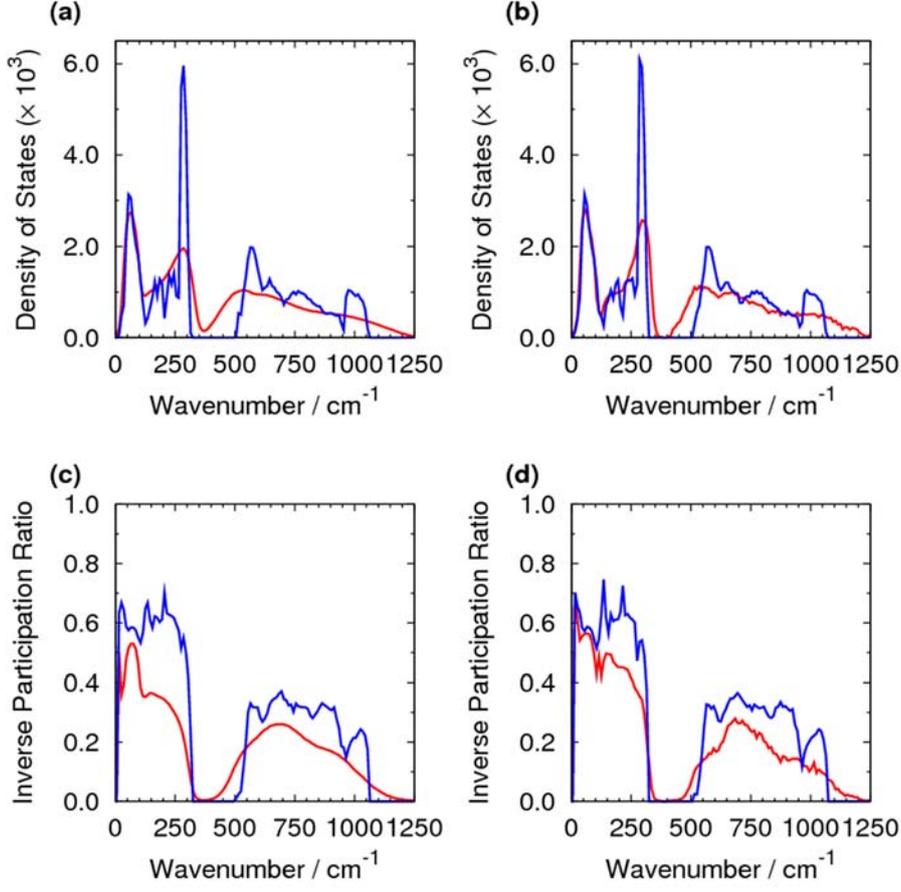

**FIG. 1.** Density of states of (a) water (red) and ice Ih (blue) at 250 K and of (b) amorphous water (red) and ice Ih (blue) at 130 K. The scaled inverse participation ratios of (c) water (red) and ice Ih (blue) at 250 K and of (d) amorphous water (red) and ice Ih (blue) at 130 K.

**Figures 2 (a)** shows the DOSs of INMs of (amorphous) liquid water at several temperatures. As seen in Fig. 2 (a), with decreasing temperature, the peaks at ~100 cm$^{-1}$ and ~400 cm$^{-1}$ decrease, whereas the peak at ~300 cm$^{-1}$ and the shoulder at ~1000 cm$^{-1}$ increase. In addition, the number of imaginary frequency modes presented in the negative frequency region decreases with lowering temperature. There exist imaginary frequency modes even below $T_g$, as known in previous studies.[14-17] $T_g$ of water was estimated to be ~135 K in our previous study. Furthermore, note that all the imaginary frequency modes are not always involved in barrier crossing dynamics.[15-17]

The scaled IPRs defined in Eq. (2) calculated with INMs are shown in **Fig. 2 (b)**. With decreasing temperature, the peak between 100 cm$^{-1}$ and 300 cm$^{-1}$ mainly due to the HB stretching mode shows a remarkable increase in delocalization, though the increase of delocalization of the other intermolecular vibration motions is marginal. Contrast to the real frequency modes, the imaginary frequency modes are localized, i.e., the decrease of IPR, with lowering temperature [inset of Fig. 2 (b)].

The inset of **Fig. 2 (c)** shows the temperature dependence of the fraction of imaginary frequency modes. When the fraction of imaginary frequency modes above 205 K is extrapolated to low temperatures, the extrapolated curve is zero at ~190 K. This sharp decrease down to ~ 190 K is due to the almost disappearance in the HDL contribution, which is also found in the temperature dependences of the structural, dynamic, and thermodynamic properties, e.g., the fraction of HB defect molecules, relaxation times, and isobaric specific heat.[9] Below 190 K, the fraction of imaginary frequency modes decreases gradually. The extrapolation of the fraction of HB defect molecules between 150 K and 190 K is zero at ~130 K, which is below the glass transition temperature of water determined in our previous work.

It has already been reported that all unstable modes are localized below the glass transition temperatures in model systems.[15] Therefore, we examined the fraction of imaginary frequency modes with larger IPR than a critical value [Fig. 2 (c)]. The critical IPR value used here is 0.18 based on the result given in Fig. 2 (b). The entropy difference between water and ice, $\Delta S$, is also superimposed in the figure. Interestingly, the temperature dependence of the fraction of extended imaginary modes is in good agreement with that of $\Delta S$ [see Fig. 8 (b)].

**Figure 2 (d)** shows the frequency-scaled INM DOSs based on the discussion by Grigera et al.[18] The frequency-scaled DOSs at 150 K and 200 K are rather linear with respect to frequency, though the finite-size peaks are seen at ~15 cm$^{-1}$ and 22 cm$^{-1}$. Since the excess DOSs at low frequencies decrease with decreasing temperature, the frequency-scaled DOS at low frequencies, for example below 10 cm$^{-1}$, also decreases due to the decrease of excess DOSs at low frequencies with decreasing temperature. As a result, the result shows that the frequency-scaled INM DOS senses the transition between minima-dominated and saddle-point dominated phase.

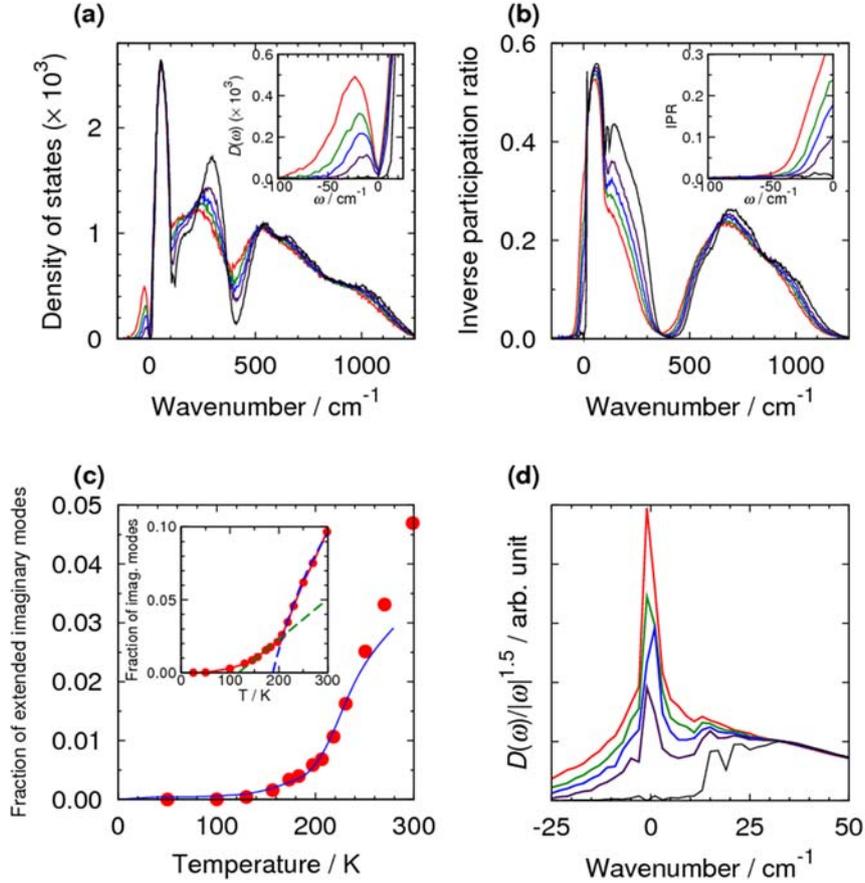

**FIG. 2.** (a) Density of states at 200 (red), 150 (green), 130 (blue), 100 (violet), and 50 K (black). Inset shows the extended figure of DOSs below 25 cm$^{-1}$. (b) Inverse participation rations (IPRs) at 200 (red), 150 (green), 130 (blue), 100 (violet), and 50 K (black). Inset shows the extended figure of IPRs for imaginary frequencies. (c) Fraction of extended imaginary frequency modes (red points) of water and scaled entropy difference between water and ice (blue). A mode whose IPR is larger than 0.18 is considered as an extended mode. Inset shows the fraction of all modes with imaginary frequency. Blue and green dashed lines are fitted above 205 K and between 150 and 190 K, respectively. (d) Frequency-scaled DOSs calculated with INM at 200 (red), 150 (green), 130 (blue), 100 (violet), and 50 K (black).

## B. Calculation of quantum-corrected specific heat and entropy

Thermodynamic properties and the Kauzmann temperature depend on the entropy, which in tur is determined by the specific heat required with quantitative accuracy over a wide range of temperatures. For most liquids, the quantum effects on the specific heat become relevant only at very low temperature, such as below 100 K. However, for water, in both amorphous and crystalline states, quantum effects are found to be important at much higher temperatures, because of the presence of the high frequency intermolecular motions. The calculation of specific heat over a wide temperature range is non-trivial; for example, the specific heat converges to a non-zero value classical mechanically, whereas it vanishes at 0 K quantum mechanically.

Here, we develop and implement an approximate but quantitatively reliable scheme to calculate the specific heat over an entire temperature range by correcting the data obtained from classical MD simulations. Note that the scheme should capture the following key behaviors of the specific heat of water; (i) the asymptotic approach to the classical limit at high temperatures, (ii) the pronounced increase found between 200 K and 240 K due to the influence of large local-density fluctuation between HDL and LDL forms of water, and (iii) the approach to zero value when the temperature goes to 0 K.

Our calculation of the quantum specific heat consists of two parts. First, we use the elegant expression of Lebowitz, Percus, and Verlet to obtain the classical specific heat. In the second part, we employ a harmonic approximation to obtain the correction due to the quantum effects. Subsequently, we extend the scheme to obtain the frequency-dependent quantum specific heat.

The expression of classical specific heat in the microcanonical ensemble is expressed as[19]

$$C^{(C)} = R\left(\frac{1}{N_f/2} - K(0)\right)^{-1}. \qquad (3)$$

Here, $R$ is the gas constant, $N_f$ is the number of degrees of freedom per molecule, i.e., six for a rigid water molecule, and $K(0)$ is the value at $t=0$ of the time correlation function (TCF) of kinetic energy fluctuation defined by

$$K(t) = \frac{N}{T^2} \langle \delta T(0) \delta T(t) \rangle, \tag{4}$$

where $T$ and $\delta T$ are the average temperature and the fluctuation of instantaneous kinetic energy of the system with $N$ molecules, respectively. Note that Eq. (3) does not possess the correct temperature dependence at low temperatures. Instead, it converges to the Dulong-Petit law, i.e., $C^{(C)} = N_f R$.

While the value of specific heat is known both in the high temperature and low temperature limits, *it is not known at the intermediate temperature range which of great importance in the present problem because we calculate the temperature dependence of entropy from the temperature dependent specific heat.* In this study, we assume that $C^{(Q)}$ consists of the classical mechanical specific heat and the correction term,

$$C^{(Q)} = C^{(C)} + \Delta C. \tag{5}$$

We further assume that the correction term, $\Delta C$, is given by the difference between quantum and classical mechanical specific heats calculated with a harmonic approximation,

$$\Delta C = C^{(Q)} - C^{(C)}$$

$$\approx C^{(Q)}_{\text{harmo}} - C^{(C)}_{\text{harmo}} = k_B \int_0^\infty \left\{ (\beta \hbar \omega)^2 \frac{e^{\beta \hbar \omega}}{(e^{\beta \hbar \omega} - 1)^2} - 1 \right\} g(\omega) d\omega, \tag{6}$$

where $k_B$ is the Boltzmann constant, $\beta$ is the reciprocal temperature, $g(\omega)$ is the density of states of normal modes. The first and second terms in Eq. (6) correspond to the quantum and classical specific heats in harmonic approximation. The correction term approaches zero in the high temperature limit, whereas it approaches the negative value of the classical specific heat to ensure the correct behavior in the low temperature limit.

An alternative representation of quantum mechanical specific heat, $C^{(Q)}$, can be obtained if we introduce the following ansatz expressed in a similar form to Eq. (3),

$$C^{(Q)} = R \left( \frac{1}{N^{(Q)}_{\text{eff}}(T)/2} - K^{(Q)}(0) \right)^{-1}. \tag{7}$$

Here $N_{eff}^{(Q)}(T)$ is *a temperature-dependent quantum mechanical effective number of degrees of freedom per molecule* and $K^{(Q)}(t)$ is the Kubo-transformed TCF of kinetic energy fluctuation. Recall that, using the Kubo-transformed TCFs, transport coefficients in quantum mechanics can be expressed in the same way as those in classical mechanics.[20, 21] A Kubo-transformed TCF is a real function, and is approximated by the corresponding classical TCF within the first-order of $\hbar$.

Although $N_{eff}^{(Q)}(T)$ cannot be determined by *a priori* manner, once $C^{(Q)}$ is obtained, we can calculate it by the following equation,

$$N_{eff}^{(Q)}(T) = 2\left(\frac{R}{C^{(Q)}} + K^{(Q)}(0)\right)^{-1}. \tag{8}$$

We now use the linear response theory again to obtain the frequency-dependent specific heat. Note that complex susceptibilities can reveal the dynamic origin of corresponding static susceptibilities. As shown by Grest and Nagel, the classical complex, i.e., frequency-dependent, specific heat, $\tilde{C}^{(C)}(\omega)$, is expressed as[22]

$$\tilde{C}^{(C)}(\omega) = R\left(\frac{1}{N_f/2} + \tilde{\dot{K}}(\omega)\right)^{-1}, \tag{9}$$

where $\tilde{\dot{K}}(\omega)$ is the Fourier-Laplace transform of the time-derivative of $K(t)$. Based on the same procedure for the classical complex specific heat, we use the following quantum mechanical complex specific heat corresponding to Eq. (9),

$$\tilde{C}^{(Q)}(\omega) = R\left(\frac{1}{N_{eff}^{(Q)}(T)/2} + \tilde{\dot{K}}^{(Q)}(\omega)\right)^{-1}, \tag{10}$$

where we should note that $C^{(Q)}$ and $\tilde{C}^{(Q)}(\omega)$ show the correct temperature dependence.

The equilibrium entropy now can be calculated by using the quantum mechanical specific heat,

$$S = \int_0^T \frac{C^{(Q)}}{T'} dT', \qquad (11)$$

and we can also introduce the complex entropy by using the complex specific heat,

$$\tilde{S}(\omega) = \int_0^T \frac{\tilde{C}^{(Q)}(\omega)}{T'} dT'. \qquad (12)$$

By using the quantum specific heat calculated from the classical data, Eq. (10), we can successfully decompose into the configurational and vibrational components; such a systematic decomposition has not been performed previously. In the present analysis, based on our previous work,[23] a contribution with frequency higher than 20 cm$^{-1}$ in the kinetic energy fluctuation is assigned to the vibrational component, whereas that lower than the threshold frequency is to the configurational component. Note that the specific heat of ice arises only from the vibrational component by using the threshold frequency. Thus, the kinetic energy fluctuation is expressed by the sum of configurational and vibrational contributions, that is,

$$\begin{aligned} K^{(Q)}(0) &= -\tilde{K}^{(Q)\prime}(0) \\ &= \frac{2}{\pi} \int_0^\infty \tilde{K}^{(Q)\prime\prime}(\omega) d\omega \\ &\equiv K_{conf}^{(Q)} + K_{vib}^{(Q)}, \end{aligned} \qquad (13)$$

where the first and second terms represent the configurational and vibrational contributions, respectively. By inserting Eq. (13) into Eq. (7), we have the specific heat expressed by,

$$\begin{aligned} C^{(Q)} &= (N_{eff}^{(Q)} R + N_{eff}^{(Q)} C^{(Q)} K_{vib}^{(Q)} + N_{eff}^{(Q)} C^{(Q)} K_{conf}^{(Q)})/2, \\ &\equiv C_{vib}^{(Q)} + C_{conf}^{(Q)}. \end{aligned} \qquad (14)$$

In Eq. (14), the term of $N_{eff}^{(Q)} R / 2$ is included in the vibrational component. Corresponding to Eq. (14), the entropy is also divided into two contributions,

$$S = S_{vib} + S_{conf}. \qquad (15)$$

## C. Kinetic energy fluctuation TCFs and relaxation times of water and ice

**Figure 3 (a)** shows the TCFs of kinetic energy fluctuation, eq. (4), of water at 250, 230, 220,

205, and 180 K. The contribution of the intermolecular vibrational motions to the energy fluctuation is seen as oscillatory motions in the sub-pico-second time scale. On the other hand, the configurational contribution is found in the slower time scale and its time dependence can be expressed by a stretched exponential function, with a relaxation time that grows rapidly as temperature is lowered. On the other hand, only the vibrational motions are found and the configurational contribution is absent in the TCFs of ice [**Fig. 3 (b)**].

The temperature dependence of relaxation time is presented in **Fig. 3 (c)**. The relaxation time shows a non-Arrhenius temperature dependence. The temperature dependence of relaxation time consists of three branches, i.e., the presence of two dynamical transitions, (i) above 220 K, (ii) between 220 K and 190 K, and (iii) below 190 K. The first dynamical transition at 220 K corresponds to the large fluctuation between HDL and LDL, whereas the molecules contributing to HDL significantly decreases below the second dynamical transition at 190K. It is found that each of the branches can be described by an individual Vogel-Fulcher-Tammann function with a different fragility parameter.[9]

The fragility of the second branch between 190 K and 220 K is weaker than that of the first branch above 220 K, reflecting the fact that the dominant contribution changes from HDL to LDL at 220 K. Along the lowest, the third branch, the number of three- and five-coordinated molecules decreases with lowering temperature and the structure, dynamics, and thermodynamic properties are governed by LDL. As a result, the supercooled state below 190 K has the weakest fragility among the three branches. This branch leads to low-density amorphous (LDA) state with lowering temperature. By extrapolating the relaxation times in the third branch, the fitted relaxation time becomes larger than 100 s at ~135 K.

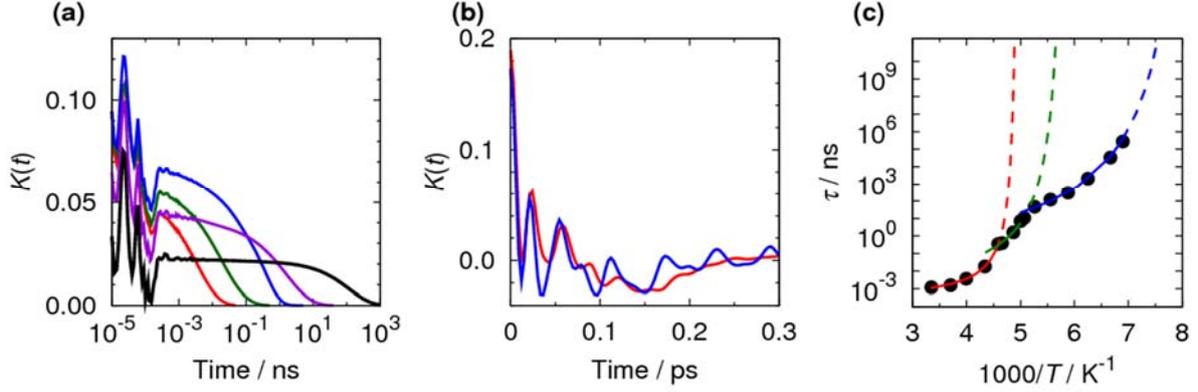

**FIG. 3.** Time correlation functions (TCFs) of kinetic energy fluctuation (a) at 250 (red), 230 (green), 220 (blue), 205 (violet), and 180 K (black) of water and (b) at 250 (red) and 100 K (blue) of ice, respectively. (c) Relaxation times of kinetic energy fluctuations. Red, green, and blue lines show three regimes of Vogel-Fulcher-Tammann (VFT) temperature dependences exhibited in three different temperature domains. For all the three domains, the relaxation time can be fitted by $\exp(DT_0/(T-T_0))$. The VFT temperature, $T_0$, and the fragility index, $D$, are 200, 170, and 120 K, and 0.58, 1.43, and 3.17, for the three regions (red, green, and blue curves). The estimated glass transition time is ~135 K form the fitted relaxation time in the third branch below ~190 K.

### D. Specific heats of water and ice

**Figure 4 (a)** shows the calculated specific heats of liquid water and ice Ih. We must stress here that *the present theoretical quantum-corrected $C_P$'s of liquid and ice are in quantitative agreement with the experimental results*[24, 25] and existing *theoretical results based on path-integral MD simulations*,[26] indicating that the harmonic correction is indeed a good approximation for the specific heats of liquid water and crystaline ice. It is important to note that the agreement is nearly perfect both at high temperature liquid and low temperature ice. This shows that our approach based on combining the Lebowitz-Percus-Verlet expression with the correction term from the harmonic approximation is a valid approach.

Figure 4 (a) shows a huge quantum effect on specific heat in both liquid water and ice. The quantum correction yields a smoothly temperature-dependent specific heat between high and low temperature limits, whereas the classical specific heat converges to the constant value of $6R$ expected from the Dulong-Petit law for a rigid water molecule. The difference between the classical and quantum mechanical specific heats becomes increasingly more pronounced as temperature is lowered and the correct quantum value is about half of the wrong classical value at ~150 K. The strong influence of quantum effects in lowering the specific heat from the classical values is important in determining the relative thermodynamic stabilities of the phases involved and $T_K$ of water glass transition (see below).

The sharp rise in the specific heat of liquid water down to ~230 K is found in **Figs. 4 (a)** and **4 (b)**. This rise is attributed to the large local-density fluctuation induced by the crossover between HDL and LDL.[27-35] As seen in Fig. 4 (b), both the vibrational and configurational contributions increase with decreasing temperature by ~230 K. The increase of the vibrational contribution due to the change of intermolecular vibrational motion induced by the mixing of HDL and LDL is larger than that of the configurational contribution. *This is partly the reason that the glass transition temperature of water is relatively lower than the melting temperature, i.e., $T_g/T_m$~1/2, different from the most of glass-forming liquids with $T_g/T_m$~2/3*. Below ~230 K, the specific heat of liquid water rapidly decreases. In particular, the configurational specific heat is very small below 190 K and almost vanishes at 130 K.

The specific heat difference between water and ice, $\Delta C_P(T)$, is shown in **Fig. 4 (c)**. Due to the increase of $C_P$ of liquid water, $\Delta C_P$ also increases at ~220 K. Note that the configurational contribution is absent in ice Ih [inset of Fig. 4 (c)]. Figure 4 (c) shows that the vibrational contribution dominants the $\Delta C_P$ in the entire range of temperature; the fraction of the vibrational contribution is ~70% down to ~220 K, and increases with further decreasing temperature, especially below ~190 K, due to the transition to LDL.

**Figure 4 (d)** presents another important result; the temperature dependence of the quantum-corrected numbers of effective degrees of freedom, $N_{eff}^{(Q)}$, of water and ice. In a rigid classical model, the number of degrees of freedom per molecule is 6. This is only valid classically. In our present approximation $N_{eff}^{(Q)}$ consistently depends on temperature as the proper temperature dependence of quantum-corrected specific heat; for example, $N_{eff}^{(Q)}$ of liquid water is ~5.6 at 300 K. This is because full thermal excitation is not achieved at this temperature because of the presence of motions with higher frequency than 300 K corresponding to ~200 cm$^{-1}$, e.g., the librational motion [for example, Fig. 1 (a)]. In water, $N_{eff}^{(Q)}$ starts to decrease below ~220 K, especially sharp decrease down to ~180 K, and eventually goes to zero at 0 K where none of the degrees of freedom is excited.

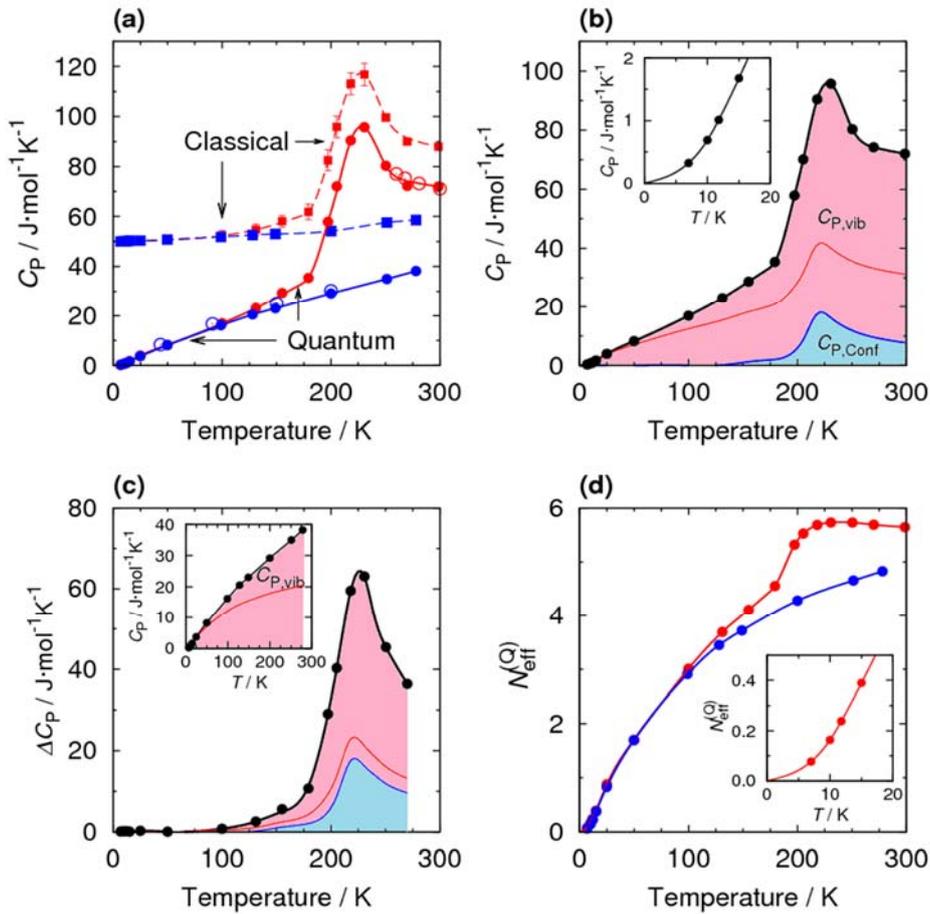

**FIG. 4.** (a) Specific heats of liquid water (red) and ice Ih (blue). Solid and dashed lines are the quantum and classical specific heats of water and ice. Red and blue open circles are specific heats of water and ice calculated from path integral simulations.[26] (b) Decomposition into the configurational (blue) and vibrational (red) contributions of the specific heat of water (black). The contribution of the numbers of effective degrees of freedom, $N_{eff}^{(Q)} R/2$, which is defined

as a part of the vibrational contribution, is shown as the region between blue and red lines. Inset shows the total specific heat below 20 K. (c) Decomposition into the configurational (blue) and vibrational (red) contributions of the difference specific heat between liquid water and ice Ih (black). The contribution of the numbers of effective degrees of freedom is shown as the region between blue and red lines. The inset shows the specific heat of quantum-corrected specific heat of ice Ih. Below 12 K, the temperature dependence of specific heat of ice is well approximated by Debye-Sommerfeld equation, $\gamma T + \alpha T^3$. The Debye temperature of ice is evaluated as 164 K. (d) Temperature dependence of quantum-corrected numbers of effective degrees of freedom of water (red) and ice Ih (blue). The inset shows the result of water below 20 K, showing the approach to zero.

### E. Entropies of water and ice

With the calculated specific hear data, we proceeded to calculate the temperature dependent entropies of liquid and crystalline water. Here, we should note the presence of residual entropy in ice, that is the extra entropy due to the multiple configurations with HBs between adjacent oxygen atoms. The experimental estimate of this contribution for ice is 3.54 J/mol K[36] and the Pauling's estimate was $k_B \ln(3/2) = 3.47$ J/mol K.[37] Since the fraction of non-four-coordinated water molecules is less than 1% in amorphous water at 130 K, and decreases with lowering temperature, this residual entropy of amorphous water is expected to be close to that of ice. As a result, hereafter the residual entropy is not explicitly considered in this study.

In **Fig. 5 (a)** we plot the temperature-dependent entropies of liquid water and ice Ih. Note that the entropies are calculated from the quantum specific heats. Corresponding to a gradual increase of $\Delta C_P$ between liquid water and ice above ~130 K, the entropy of liquid water becomes gradually larger than that of ice above the temperature. In addition, the entropy of liquid water sharply increases between 190 K and 230 K arising from the sharp increase in $C_P$ of liquid water in that range of temperature, due to the presence of large-scale density fluctuations in the liquid.

**Figure 5 (b)** shows the calculated temperature dependence of the *entropy difference* between liquid and crystalline water, $\Delta S$. It is found that the decrease in $\Delta S$ occurs in two stages: the first sharp variation occurs down to ~190 K and then the second decrease is slow and gradual. **Figure 5 (c)** shows the decomposition of $\Delta S$ into two apparently disjoint entropy contributions, the configurational and vibrational components. Here, as mentioned in subsection B, the configurational and vibrational components are calculated using the frequency-dependent entropies of liquid water and ice [details are seen in Fig. 8 (b)]. The early part of the decrease in $\Delta S$ with decrease in temperature is due to the decrease in both the configurational and the vibrational contributions. *If we extrapolate this early part to obtain the temperature of zero entropy difference between liquid and ice, we would get $T_K$~190 K.* Here, we note that $T_K$ is by definition a value obtained by extrapolation. The extrapolation of the second decrease that originates primarily from *the vibrational motions* [Fig. 5 (c)], however, leads to $T_K$ being equal to ~130 K. Below 130 K, the entropy difference between amorphous water and ice is indeed very small, less than 0.5 J/cal$^{-1}$K$^{-1}$. *Thus, the estimated value of $T_K$ is slightly below our estimated $T_g$, 135 K, where our fit to the calculated relaxation time becomes ~100 s* [Fig. 3 (b)]. *Here, the experimental $T_g$ of LDA is 136 K.*[38]

The relationship between the configurational entropy and the relaxation time of the self-term of intermediate scattering function is presented in **Fig. 5 (d)**. According to Adam-Gibbs relation,[6] the relaxation time $\tau_\alpha$ should exhibit a dependence on the configuration entropy as $\tau_\alpha \propto$ exp($B/TS_C$). Figure 5(d) however shows that such a dependence is no longer valid across the entire temperature range.

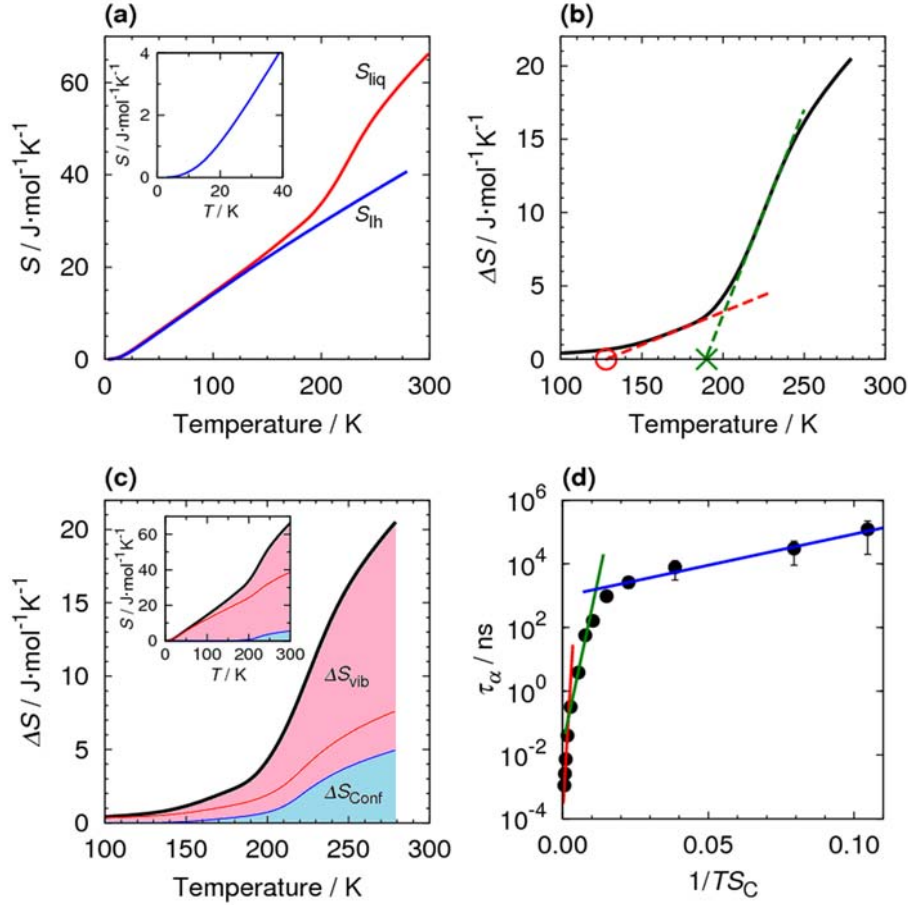

**FIG. 5.** (a) Entropies of liquid water (red) and ice Ih (blue). Inset is the entropy of ice below 50 K. (b) Entropy difference between liquid and crystal (red). Blue and green dashed lines, which are fitted above 205 K and between 150 and 190 K, cross zero at 192 K (cross) and 130 K (circle). (c) Total entropy difference between liquid water and ice Ih (black), and its configurational (blue) and vibrational (red) contributions. The contribution of the numbers of effective degrees of freedom is shown as the region between blue and red lines. (d) Relationship between the configurational entropy and the relaxation time of the self-term of intermediate scattering function. Red, green, and blue are fitted lines with exp(1/$TS_C$) above 230 K, between 230 K and 190 K, and below 190 K, respectively.

### F. Gibbs energies of water and ice

Armed with the accurate specific heats and entropies of both liquid water and crystalline ice, we calculated the Gibbs energies of the two phases (**Fig. 6**). The dashed curves in Fig. 6 are the Gibbs energy of (amorphous) liquid water and ice Ih calculated with $U + 3RT + pV - TS$. These two Gibbs energies cross at 258 K, which is in good agreement with the melting point ~

250 K calculated from the simulations of the system with coexistence of a solid-liquid interface and free energy calculation.[39]

The solid lines in **Fig. 6** show the Gibbs energy $U + (N_{eff}^{(Q)}/2)RT + pV - TS$ of water and ice. As shown in Fig. 4 (d), the number of effective degrees of freedom $N_{eff}^{(Q)}$ of liquid water becomes close to 6 above ~200 K, and thus $U + (N_{eff}^{(Q)}/2)RT + pV - TS$ approaches $U + 3RT + pV - TS$. On the other hand, $N_{eff}^{(Q)}$ of ice is still less than 5 even near 280 K, and thus free energy of ice is smaller than $U + 3RT + pV - TS$. As a result of the fact that the difference between the effective thermally excited degrees of freedom in water and ice, *the present result predicts a melting temperature at ~280 K, in fair agreement with the experimental value of 273 K*. Given that our calculations involve no adjustable parameter, this can be regarded as satisfactory. *The present result shows that quantum effects play a role even in the melting point of ice!* Note further that the free energy curve of liquid water shows a bimodal peak with a dip at ~190 K, which is the temperature of the second dynamical transition.

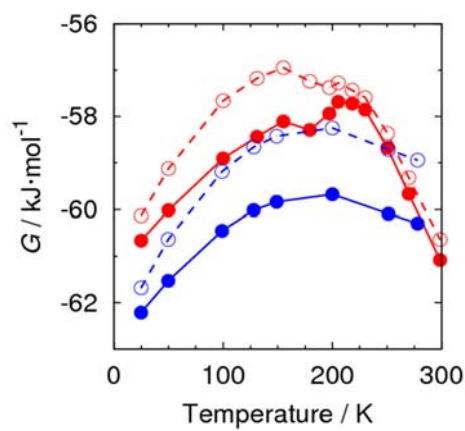

**FIG. 6.** Gibbs energies of liquid water (red) and ice Ih (blue). Dashed lines are the Gibbs energies of liquid water and ice in which the number of degrees of freedom is assumed to be 3, whereas solid lines are those with the number of effective degrees of freedom, $N_{eff}^{(Q)}$ given in Fig. 4 (d).

**G. Complex, frequency-dependent, specific heats of water and ice**

**Figures 7** show the frequency dependence of the imaginary parts of the quantum-corrected complex specific heats of water and ice at several different temperatures. The peaks arising from the intermolecular vibrational motions above 20 cm$^{-1}$ in both water and ice obviously correspond to the HB bending, HB stretching, and libration motions. *Below 20 cm$^{-1}$, however, there exists a peak at low temperatures that exhibits a remarkable temperature dependence.* This peak is attributed to the configurational fluctuations, i.e., the HB network rearrangement dynamics of liquid water. This peak becomes large when temperature approaches 220 K that is attributed to the fluctuation between HDL and LDL.[23] With decreasing temperature below 220 K, the amplitude of the configurational fluctuations becomes small rapidly, and their timescale continues to be slow. Note that such configurational fluctuations are absent in ice. This configurational component plays an important role in determining the temperature-dependent anomalies of liquid water.

Figure 7 also shows that, the imaginary part of complex specific heat, i.e., the effective density of states for the kinetic energy fluctuation, of all the intermolecular vibrational motions of liquid water is maximum at ~220 K, that is the intermolecular vibrational energy fluctuation also become large due to the large local-density fluctuations between HDL and LDL. On the other hand, that of ice monotonically decreases with decreasing temperature. The frequency dependence of specific heat offers a good marker of the collective dynamics of the system, especially of the correlated energy fluctuations in the complex landscape of the liquid.

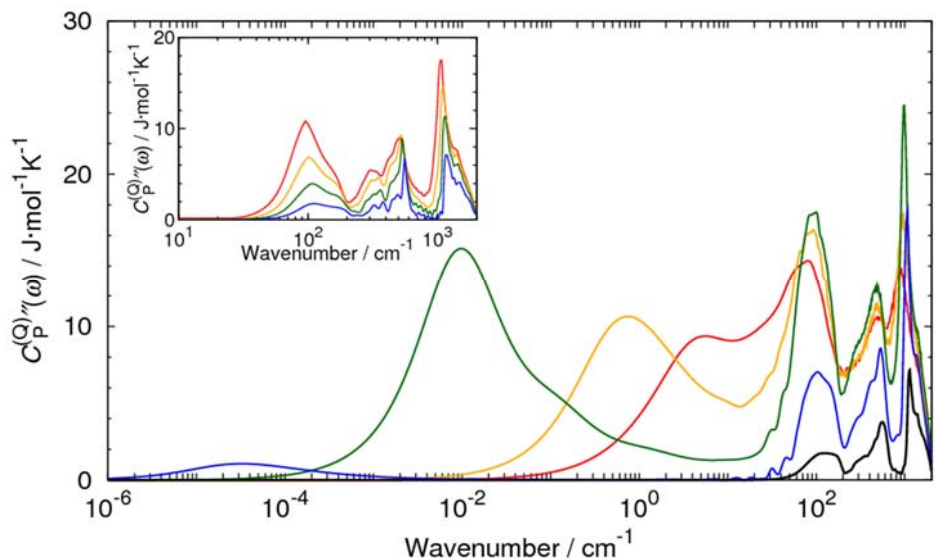

**FIG. 7.** Imaginary parts of quantum-corrected complex specific heats of water at 300 (red), 250 (orange), 220 (green), 180 (blue), and 100 (black) K. Inset shows those of Ice Ih at 250 (red), 200 (orange), 150 (green), and 100 (blue) K.

**H. Complex frequency dependent entropies of water and ice**

As already discussed, the clear separation between the vibrational and configurational contributions to the total specific heat of supercooled water allows us to define a frequency-dependent entropy which displays certain unique features as detailed below.

In **Fig. 8 (a)** we show the temperature-scaled frequency dependence of specific heat. The figure shows that the low frequency configurational contribution becomes progressively separated from the vibrational band as the temperature is lowered. Note also the progression toward lower frequency and the decrease in amplitude of the configurational band. The clear separation in time scales between the vibrational and configuration contributions to specific heat motivates a definition of frequency-dependent entropy in terms of the imaginary part of complex specific heat, eq. (12).

**Figure 8 (b)** shows the *frequency-dependent entropy, $S''(\omega)$, of liquid water* at several temperatures. At very low temperatures, below $T_g$, only the distinct peaks arising from the vibrational component are seen above ~20 cm$^{-1}$ in $S''(\omega)$, whereas no configurational component arising from the HB network structural changes is found in $S''(\omega)$. Above $T_g$, the configurational contribution emerges at very low frequencies and at ~220 K a low frequency peak from ~$10^{-4}$ cm$^{-1}$ to ~$10^{-1}$ cm$^{-1}$, which is attributed to the local-density fluctuation between HDL and LDL, is found. With further increasing temperature, the configurational entropy arising from the HB network dynamics can be seen in a wide range of frequency. The present results reveal that the shoulder at ~$10^{-2}$ cm$^{-1}$ is the remnant of the fluctuation between HDL and LDL at lower temperatures, i.e., ~220 K, whereas the peak at ~$10^{0}$ cm$^{-1}$ arises from the HB dynamics of HDL. Note that the long tail towards low frequency in $S''(\omega)$ is not present in ice.

In **Fig. 8 (c)**, we plot the difference of frequency-dependent entropy between liquid and ice at 200 K. The frequency-dependent excess entropies are found in small frequency ranges, between ~50 cm$^{-1}$ and ~80 cm$^{-1}$ and between ~700 cm$^{-1}$ and ~1000 cm$^{-1}$. There is noticeable excess entropy in the very small frequency, ~$10^{-4}$ cm$^{-1}$, due to the rare barrier crossing dynamics involving the defects evident from system trajectories. Our recent study demonstrated a clear correlation between the growth of structural relaxation time and the decrease of $H^1O^2$-defects, indicating that the $H^1O^2$-defects play a crucial role in such rare HB network structural changes in water.[9]

**Figure 8 (d)** shows the DOSs scaled by squared frequency, of four-coordinated molecules and of the HB defects, i.e., three- and five-coordinated molecules, in water and of ice, calculated from the velocity time correlation function. The excess DOSs at ~30 cm$^{-1}$ and from ~500 cm$^{-1}$ to ~700 cm$^{-1}$ correlate with the low frequency excitation due to the defects, i.e., the Boson peak, and the softening of libration motion [Figs. 1 (a) and 1 (b)]. Although suggestive,[16] yet the simultaneous disappearances of the extended unstable modes[15] and of the Boson peak is remarkable [Figs. 2 (c) and 8 (d)]. As discussed above, the temperature dependence of the fraction of extended imaginary modes exhibits a *crossover which is similar to the crossover observed in ΔS* [Fig. 2(c)]. The fraction of extended imaginary frequency modes may thus serve as an order parameter of the transition between minima-dominated and saddle-point dominated phases, as envisioned in several theories based on the energy landscape picture of the glassy liquids.[20,21] Interestingly, this transition temperature is close to our estimate of the Kauzmann temperature. A quantitative relation among Boson peak, intensity of unstable modes, and relaxation spectrum of glassy liquids is yet to be developed.

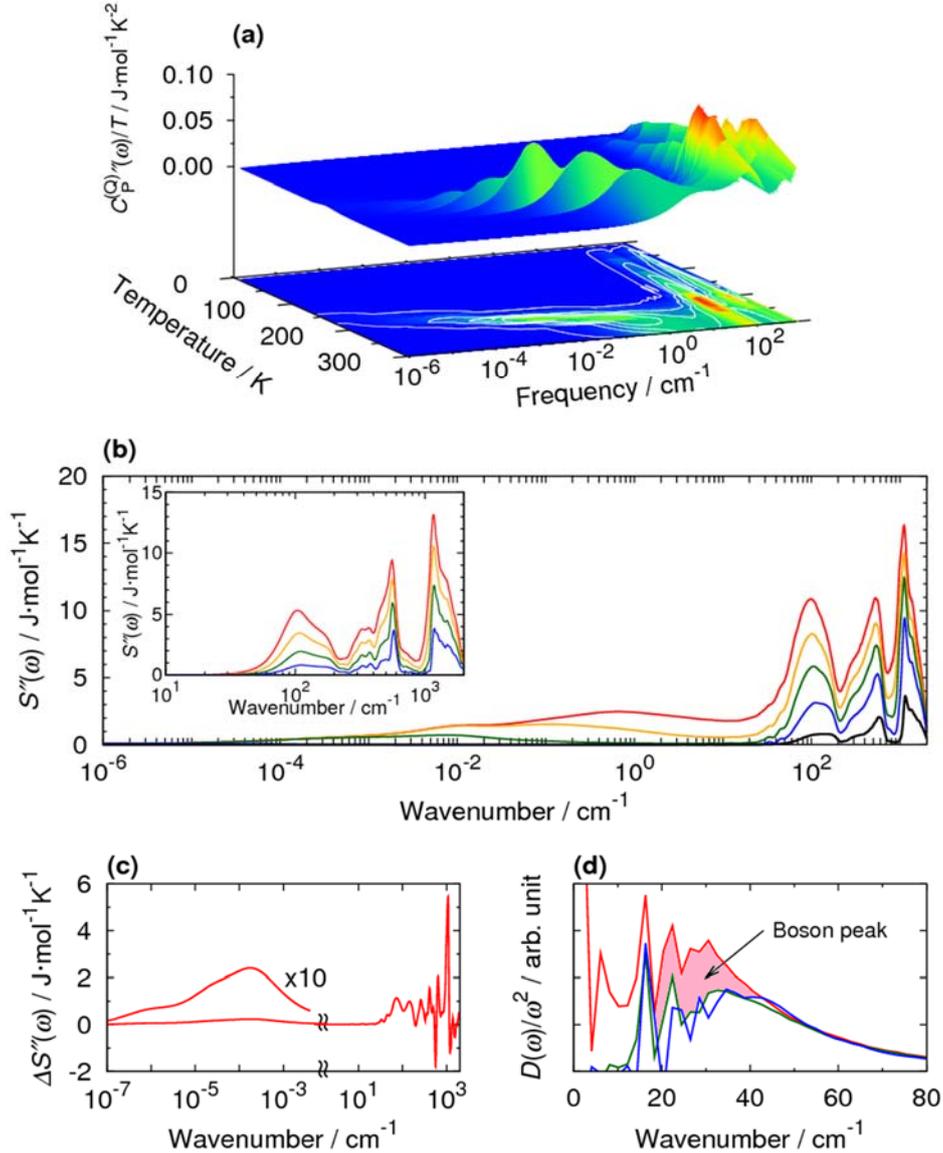

**FIG. 8.** (a) Temperature-scaled imaginary part of complex specific heat of water. (b) Frequency-dependent entropies of (liquid, supercooled, and amorphous) water at 300 (red), 250 (orange), 220 (green), 180 (blue), and 100 (black) K. Note the clear separation between the vibrational and configurational contributions below 250 K and the configurational entropy disappears at 100 K. Inset shows those at 250 (red), 200 (orange), 150 (green), and 100 (blue) K of ice Ih. (c) Difference in frequency-dependent entropy between liquid water and ice Ih at 200 K. Note the change in scale. (d) DOSs scaled by squared frequency of four-coordinated molecules (green), HB defect molecules (red), and ice (blue), respectively. Note the excess density of states of the defect molecules at ~30 cm$^{-1}$ (region shown in red). Sharp peaks below ~20 cm$^{-1}$ are due to the finite size effect found in the previous study.[40]

**I. Translational and orientational Lindemann ratios of water and ice**

We examined the translational and orientational Lindemann ratios (LRs) calculated from

$$\delta^{trans} = \frac{\sqrt{\langle \delta r^2 \rangle}}{\langle r_{CC} \rangle}, \tag{17}$$

and

$$\delta^{orien} = \frac{\sqrt{\langle (\delta \mathbf{e})^2 \rangle}}{\pi/2}. \tag{18}$$

Here $\langle r_{CC} \rangle$, $\langle \delta r^2 \rangle$, and $\langle (\delta \mathbf{e})^2 \rangle$ are the average distance between the nearest neighbor center of masses (COMs), mean square fluctuation of COM, and mean square fluctuation of dipolar vector of water molecules.

In **Figs. 9 (a)** and **9 (b)**, we present the temperature dependence of the translational and orientational LRs of liquid water and ice. Note that one can easily detect the onset temperature of the transition to the mobile liquid state, ~130 K. Both the two different LRs of water reveal the presence of the same four regions, i.e., (i) below 130 K where the liquid is frozen, (ii) between 130 K and 190 K where we see the first indications of melting, (iii) between 190 K and 220 K which is clearly a region of rapid change, and (iv) above 220 K where we see the saturation when HDL is the dominant liquid state. The three regions, (ii) – (iv), correspond to the three branches in the temperature dependence of relaxation times of the kinetic energy fluctuation [Fig. 3 (b)] and the self-term of intermediate scattering function.[9]

The temperature dependence of the Lindemann ratio is routinely used to diagnose the melting of a solid.[41] As shown in Figs. 9 (a) and 9 (b), both the translational and orientational LRs show the melting process to start ~135 K. More importantly, *the temperature dependence of the rise* *of LR for both the translational and orientational displacements track the variation of the total entropy difference between glassy water and ice* [**Fig. 9 (c)**]. Such a relation between entropy and LR may be rationalized from the self-consistent phonon theory of glass transition.[42, 43]

Although connections among defects, Boson peak, entropy, relaxation rates, and glass transition have long been discussed in the literature, the nature of defects in glassy liquids is often left unclear. In the case of deeply supercooled water, the definition and identification of defects are unambiguous, because four-coordinated water molecules form a continuous network that spans the system, while three- and five-coordinated molecules appear as defects. We established that the excess DOSs, e.g., the Boson peak, could be associated with three- and five-coordinated defects that catalyze relaxations at low temperatures. These are reflected in the frequency-dependent specific heat and frequency-dependent entropy. The excess entropy originates not just from softening of modes but also from motion of the defects in the network.

Our work could establish an important correlation with the work of Parisi and co-workers by examining the fraction of extended unstable modes.[20] *Interestingly, the temperature dependence of the fraction of extended unstable modes is in good agreement with that of the entropy difference between liquid water and crystalline ice.* The fraction of extended unstable modes is also found to be correlated with the amplitude of the Boson peak. Both are found to disappear below the glass transition temperature. The present result suggests that these quantities give a signature for the transition between minima-dominated and saddle-point dominated phases.[20]

Stability of a glass to crystallization is an important current issue. Here the coupling between defects and the intermolecular vibrational modes play critical role both in the thermodynamics and dynamics of glassy liquids, including the determination of glass transition temperature. These modes are largely absent in atomistic models (such as Kob–Andersen) of glass transition phenomenon but are of paramount importance in the vitrification of water, and may be of other molecular (and network) glasses.

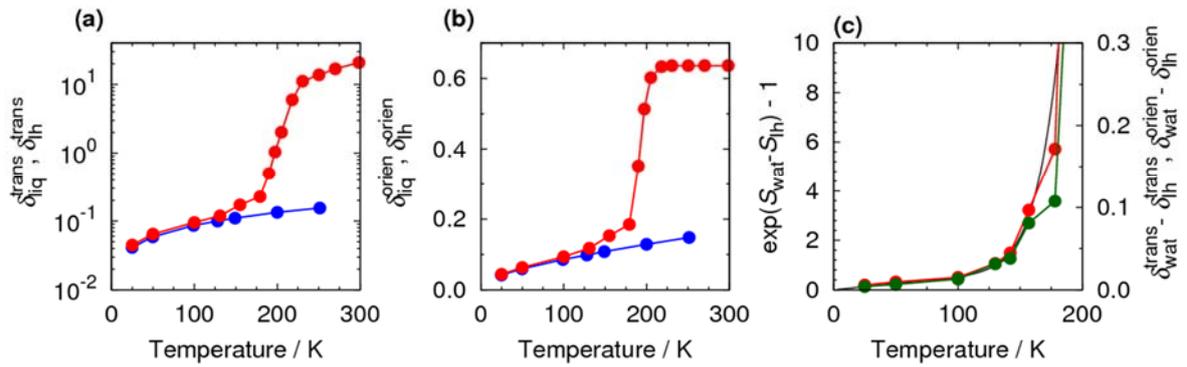

**FIG. 9.** (a) Translational and (b) orientational Lindemann ratios of liquid water (red) and ice Ih (blue). (c) Comparison between the entropy difference between liquid water and ice Ih (black) and the Lindemann ratio differences between water and ice. Red and green dashed lines are the translational and orientational difference Lindemann ratios.

**J. Separation dependence of orientational correlations of amorphous water and ice ih**

Amorphous ice that forms below $T_g$ remains locally ice-like but the nearly perfect tetrahedral order around an individual water molecule (seen in ice) fails to propagate as the unit cells undergo rotation with respect to each other. As a result, they get out of phase even at small separations. This is reflected in **Fig. 10 (a)** where we have plotted the separation distance dependence of the distribution of relative orientation of the individual water molecules. Figure 10 (a) clearly shows the absence of this propagation of orientational correlation in amorphous water. Somewhat surprisingly, even at low temperatures, the orientational distribution of the amorphous ice remains that of a disordered liquid with a very short correlation length, shortened by the distortion of tetrahedral structure and also by the presence of three- and five-coordinated defects. On the other hand, such orientational correlation is quite long ranged in ice, as can be seen from **Fig. 10 (b)**. The orientational arrangement in amorphous ice is not unique, and one glassy state just resides in one of many possible random arrangements of molecules, with partial local translational order but lacks orientational order.

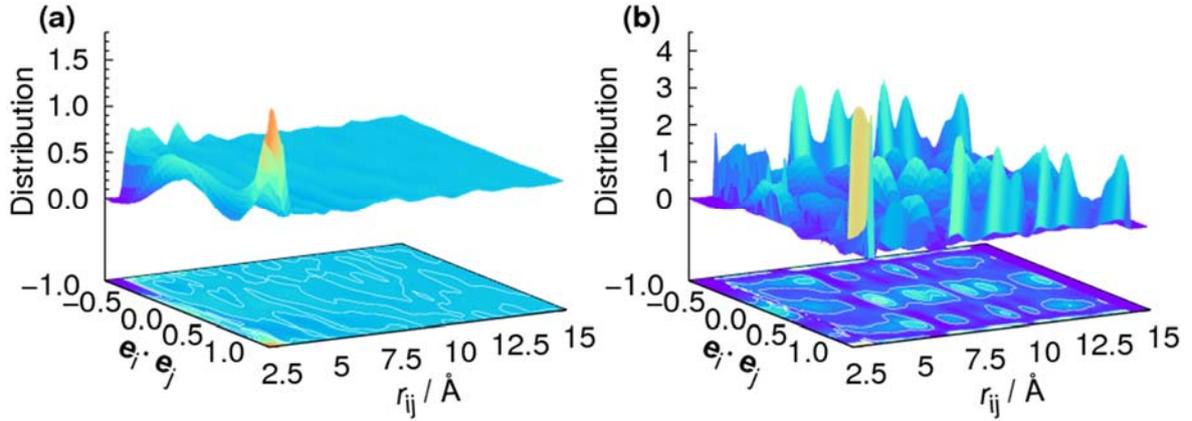

**FIG. 10.** Distance dependence of orientation correlations of (a) amorphous water and (b) ice Ih at 130 K. The long-range orientational correlation exists in crystalline ice, whereas that is almost completely lost beyond the third-hydration shell in amorphous water.

## IV. CONCLUSION

In the present study, we have attempted to develop a thermodynamic perspective of the water glass transition phenomenon. To gain this perspective, we have performed a quantitative calculation of the quantum-corrected specific heats of supercooled liquid water and ice. For the calculations we have used the seminal work by Lebowitz, Percus, and Verlet to obtain the classical specific heat and made the harmonic correction to ensure the correct temperature dependence. The specific heat so obtained was found to be in quantitative agreement with all the known experimental and computational results at all temperatures. We have also found the large quantum effects in the specific heat at low temperatures. It was revealed that the quantum correction turns out to be of crucial in determining the value of the specific heat, and of the entropy.

By applying the idea of complex heat capacity developed by Grest and Nagel to the developed quantum-corrected specific heat, we have then calculated the frequency-dependent specific heats of liquid water and ice. Because of the clear separation between the vibrational and configuration contributions to the specific heat, we have successfully distinguished the vibrational and configurational contributions. We have clarified how the configurational and vibrational motions change on approach toward the vitrification of water. Furthermore, it was revealed that the specific heat of supercooled water at low temperatures is dominated by the vibrational contribution and the specific heat difference between liquid water and ice is also derived from the vibrational contributions.

We have calculated the entropies of liquid water and ice using the quantum-corrected specific heats. The entropy so obtained has revealed a highly interesting temperature dependences of the entropy of liquid water; the rapid change between 190 K and 230 K and the slow change below 190 K. In addition, we have found a remarkable correlation between the difference entropy and the extended unstable modes. The separation between the vibrational and configurational contributions of the specific heat allowed us to define a frequency-dependent entropy. The frequency-dependent entropy has also revealed the interesting aspects of coupling between thermodynamics and dynamics (relaxation) in the low temperature liquid.

By calculating the entropies of liquid water and ice over a wide range of temperature, we have first estimated the Kauzmann temperature of water glass transition to be 130 K. The calculated $T_K$ is slightly below the glass transition temperature, 135 K, which is evaluated by the fitting of the calculated relaxation times. Note that the experimental glass transition temperature is 136 K. In addition, we have found that the configurational contribution to the entropy is very small below 190 K, where the deeply supercooled water is characterized as LDL, and thus the calculated $T_K$ of water is mainly determined by the vibrational contribution.

We have investigated the translational and orientational Lindemann ratios (LRs). It was found that the temperature dependence of both the LRs shows the presence of four temperature regions with different fluctuations towards the vitrification process of liquid water. Thus, the present study has revealed that the LRs can correctly pick up the water glass transition, and lend a valuable support to the thermodynamic and dynamic estimates of the glass transition temperature.

Finally, the separation distance dependence of orientational correlation function was found to be vastly different for ice in the amorphous state and in the crystalline state. In fact, the short correlation length came to us as a bit of surprise. We suggested that this dependence could serve as a powerful marker of the lack of order in the amorphous ice.

In the subsequent studies, we shall attempt to explore the role of defects in promoting relaxation as the amorphous ice is heated above the glass transition temperature.

**ACKNOWLEDGMENTS**

We thank Professor Iwao Ohmine for many stimulating discussions and criticisms over the course of this work. The present study was supported by JSPS KAKENHI Grant No. JP16H02254 for S.S. and by JC Bose Fellowship (DST, India) and DST-Physical Chemistry and DST-Nano Mission for B. B and the Indo (DST)-Japan (JSPS) bilateral collaboration program for S.S. and B.B. The calculations were carried out by using the supercomputers at Research Center for Computational Science in Okazaki.